\documentclass[a4paper,showpacs, 12pt]{revtex4}
\usepackage{graphicx}
\usepackage{amsfonts}
\usepackage{amssymb}
\usepackage{amsmath}

\begin{document}

\title{The globule-coil transition in a mean field approach}
\author{Erik Van der Straeten}
\author{Jan Naudts}
\affiliation{Departement Fysica, Universiteit Antwerpen, Groenenborgerlaan 171,
2020 Antwerpen, Belgium}
\email[E-mail]{Erik.VanderStraeten@ua.ac.be,Jan.Naudts@ua.ac.be }

\begin{abstract}
In this paper we study the unfolding transition observed in polymer stretching
experiments. We use a known model and extend it with a mean-field-like
interaction. Expressions for all necessary thermodynamic variables as a function
of the model parameters are derived in closed form. To obtain the exact
force-extension relation at constant temperature, we invert these relations
numerically. Below a critical temperature the model exhibits a sharp first order
phase transition. 
\end{abstract}


\maketitle

\section{Introduction}
The unfolding transition of polymers is observed in different types of
experiments. The unfolding is a single- or multi-step process, depending on the
experimental conditions.

In force-clamp experiments, one is able to study the mechanical unfolding of
polymers at relatively constant force. In such experiments, one applies a sudden
force (the time to complete the force step is of the order of $10$ms
\cite{ref1}). With feedback techniques the force is then kept constant. Both
experiments \cite{ref1} and numerical simulations \cite{ref2} show that
a certain protein (called ubiquitin) unfolds in a single step, while an other
protein (called integrin) unfolds in multiple steps \cite{ref2}. In
\cite{ref3,ref4}, force-extension relations of single DNA molecules are
measured in the fixed stretch ensemble. In these experiments, one keeps the
extension of the DNA molecule relatively constant, and the average applied force
is measured. Depending on the solvent conditions force plateaus or stick-release
patterns are observed.

The unfolding of polymers is also studied under more dynamical conditions. The
two most common are the mechanical stretching at a force that increases linearly
with time (force-ramp) \cite{ref5} and the deformation of polymers in a flow
\cite{ref6,ref7,ref8}. Numerical simulations show that under these
conditions, the unfolding is mostly a multi-step process. 

The unfolding transition from compact globule to stretched coil was already
theoretically predicted 15 years ago \cite{ref9}, based on heuristic
arguments. Nowadays, the theoretical study of this transition is dominated by
numerical simulations. Instead of performing numerical simulations, we study a
simplified model that can be solved in closed form almost completely. Despite
the simplicity of the model, it still exhibits a first order phase transition
for every chain length at low enough temperatures. For this reason, the model is
also interesting on a purely theoretical ground. 

In \cite{ref10,ref11} we introduce a model to describe polymers. In the
present paper, we start with shortly reviewing the results obtained in these
papers. We proceed with deriving expressions in closed form for the temperature
and the external force by minimizing the free energy.  These results are used to
study the unfolding transition of the polymer. The last part of the paper gives
a discussion of the results.

\section{Model}
\begin{figure}
\begin{center}
{\includegraphics[width=0.48\textwidth]{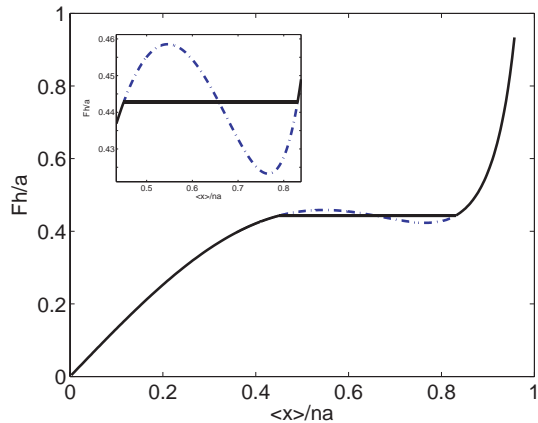}}
\caption{\label{fig:forex}(color online). Plot of the force-extension relation
at constant temperature ($h\beta=0.7$). The black solid line is the stable
solution. The blue dotted line corresponds with the unstable solution. The
length of the chain is $10$.}
\end{center}
\end{figure}
Our model is a one-dimensional persistent random walk with an external force.
It is described in detail in References \cite{ref10} and \cite{ref11}.
The model depends on two parameters $\epsilon$ and $\mu$, the probabilities to
go straight on when walking to the right, respectively to the left. This is not
a Markov chain since the walk remembers the direction it comes from. Two
important variables are the end-to-end distance $x$ and the number of reversals
of direction $k$ (number of kinks). In \cite{ref10}, the generating function
method is used to obtain exact analytical expressions for the average end-to-end
distance and the average number of kinks 
\begin{eqnarray}
\frac{\langle x\rangle}{na}=\frac{\epsilon-\mu}{2-\epsilon-\mu}&,&\frac{\langle
k\rangle}{n}=2\frac{(1-\epsilon)(1-\mu)}{2-\epsilon-\mu},
\end{eqnarray} 
with $a$ the lattice parameter and $n$ the total number of steps. Also the
average of the squared end-to-end distance can be calculated with the same
method (for the result, see appendix \ref{app1}). Starting from the definition
of Boltzmann, the following expression for the entropy $S$ is obtained in
\cite{ref11}
\begin{eqnarray}\label{xk}
S&=&-\frac{\langle
k\rangle}{2}\ln\frac{(1-\epsilon)(1-\mu)}{\epsilon\mu}-\frac{\langle
x\rangle}{2a}\ln\frac{\epsilon}{\mu}-\frac{n}{2}\ln\epsilon\mu.
\end{eqnarray}
(We use units for which $k_\textrm B=1$.)

\section{Thermodynamics}
\begin{figure}
\begin{center}
{\includegraphics[width=0.48\textwidth]{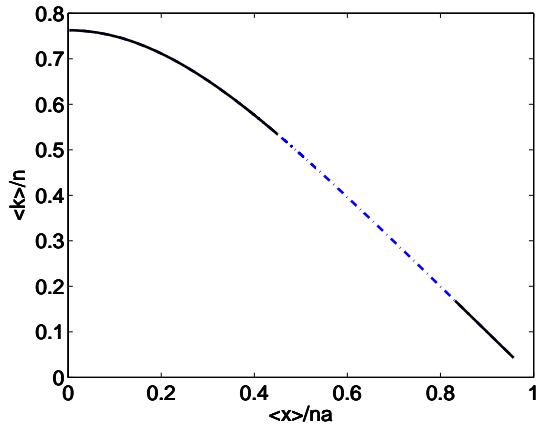}}
\caption{\label{fig:knik}(color online). Plot of the average number of kinks as
a function of the extension at constant temperature ($h\beta=0.7$). The black
solid line is the stable solution. The blue dotted line corresponds with the
unstable solution. The length of the chain is $10$.}
\end{center}
\end{figure}
The model is not yet defined in a unique way, because we did not define a
Hamiltonian. Experimentally, the ground state of the polymer is a compact
globule at vanishing force. An obvious definition is then $H=-hk$, with $h$ a
positive constant with dimensions of energy. With this choice, the probability
distribution is a Gibbs distribution, in very good approximation
\cite{ref10}. As a consequence no phase transition can occur. 

The microscopic model with $H=-hk$ does not take the excluded volume effect into
account. This is a long range effect. It is well known that
long range interactions can cause phase transitions even in finite
one-dimensional systems. So it is appropriate to take the following definition
for the Hamiltonian
\begin{eqnarray}\label{energie}
H&=&\frac{h}{a^2}\left(x-\langle x\rangle\right)^2,
\end{eqnarray}
instead of the more obvious definition mentioned above. With this definition,
the ground state is a compact globule in absence of an external force.
But also the excluded
volume effect is partly taken into account. This definition is also acceptable
form a thermodynamic point of view for two reasons. The energy is extensive. It
is well known that the thermal fluctuations of an extensive quantity (like the
end-do-end distance) are proportional to the thermal energy in classical
systems.

The Legendre transform of $S$ is the free energy $G$
\begin{eqnarray}
G&=&\inf_{\epsilon,\mu}\left\{E-F\langle x\rangle-\frac{1}{\beta}S\right\}.
\end{eqnarray}
The solution of the set of equations $\partial G/\partial\epsilon=0$, $\partial
G/\partial\mu=0$ for $\beta$ and $F$ is given in the appendix \ref{app1}. In the
thermodynamic limit this solution simplifies to
\begin{eqnarray}\label{bet_F}
\frac{a}{h}F&=&\frac{1}{2h\beta}\ln\frac{\epsilon}{\mu}-2(\epsilon-\mu)\frac{
2+\epsilon+\mu}{(2-\epsilon-\mu)^2}
\cr
h\beta&=&\frac{1}{8}(2-\epsilon-\mu)^2\ln\frac{1-\epsilon}{\epsilon}\frac{1-\mu}
{\mu}.
\end{eqnarray}
The most general case when $\epsilon\neq\mu$ corresponds with a persistent
random walk with drift. A persistent random walk is obtained with the choice
$\epsilon=\mu$. This implies $F=0$ but non-vanishing $\beta$. Also random walk
with drift is a special case. This corresponds with $\epsilon+\mu=1$ and implies
$\beta=0$ but non-vanishing $F$. Simple random walk is obtained with
$\epsilon=\mu=1/2$. In this case both $\beta$ and $F$ equal zero. 

The set of equations (\ref{bet_F}) or (\ref{bet_F_app}) has multiple solutions
for most values of $\beta$ and $F$. The stable solution is the one which
minimizes $G$.

\section{Results}
\begin{figure}
\begin{center}
{\includegraphics[width=0.48\textwidth]{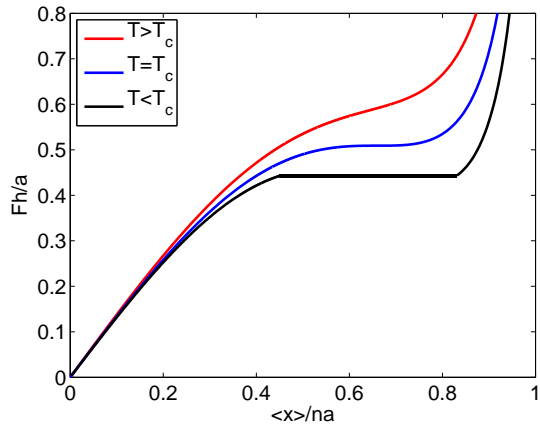}}
\caption{\label{fig:forex_3}(color online). Plot of the force-extension relation
at constant temperature, for different values of the temperature. The length of
the chain is $10$ for all three curves. The value of $h\beta$ equals $0.6$,
$0.651$ and $0.7$ for the red, blue and black curve respectively.}
\end{center}
\end{figure}
By inverting the equations (\ref{xk}, \ref{bet_F}) or (\ref{xk},
\ref{bet_F_app}) one can calculate the force-extension relation at constant
temperature for any length of the chain. This relation is the central result of
the present paper. It is shown in figure \ref{fig:forex}. The black solid line
corresponds with the stable solution (the solution which minimizes the free
energy). For small forces, the end-to-end distance grows approximately linear
with the applied constant force. At $Fh/a\approx0.44$, the polymer undergoes a
transition from a compact globule to a stretched coil. This is clearly visible
in the figures \ref{fig:forex} and \ref{fig:knik} because of the sudden increase
respectively decrease of the average end-to-end distance and average number of
kinks. After the transition, the force needed to extend the polymer further
grows in a strongly non-linear manner. 

The free energy has two minima below the critical temperature. One of these
minima corresponds with the compact globule, the other with the stretched coil.
When the free energy of these two states is equal, they coexist. Above
the critical temperature, the phase transition disappears. This can be seen in
figure \ref{fig:forex_3}, where the force-extension relation is shown for three
different values of the temperature ($T<T_c$, $T=T_c$ and $T>T_c$).

\begin{figure}
\begin{center}
{\includegraphics[width=0.48\textwidth]{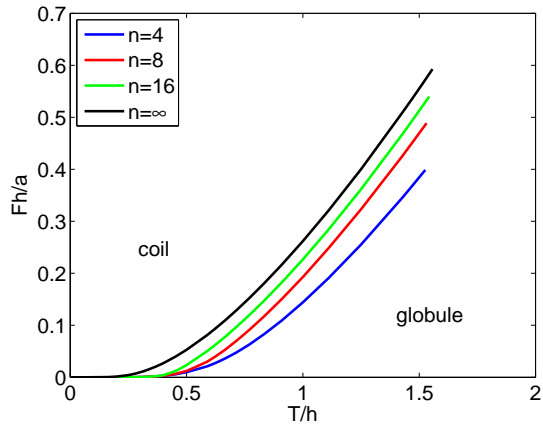}}
\caption{\label{fig:fase}(color online). The phase diagram for different values
of the chain length. The critical line moves to higher forces when the length of
the chain increases.}
\end{center}
\end{figure}

In figure \ref{fig:fase} the phase diagram is shown for different lengths of the
chain. The critical line is first order, up to the critical point. At this point
the phase transition is second order. The critical line is obtained by
numerically solving the set of equations (\ref{bet_F}) respectively
(\ref{bet_F_app}). Below the critical point, these sets have three solutions.
They correspond with two minima and one maximum of the free energy. At the
critical line we have the extra condition that the values of the two minima have
to be equal. This is also checked numerically.

The unstable regime includes two points at which $\partial F/\partial\langle
x\rangle=0$ (see figure \ref{fig:forex}). At the critical point these two points
coincide and the intermediate regime, where $\partial F/\partial\langle
x\rangle<0$, vanishes. As a consequence, the critical point can be obtained by
solving the set of equations $\partial F/\partial\langle x\rangle=0$,
$\partial^2 F/\partial\langle x\rangle^2=0$. This can be done numerically for
every length of the chain. In the thermodynamic limit one obtains following
values for the the critical temperature and critical external force
$T_c/h\approx1.557$ and $F_ch/a\approx0.5927$ 

\section{Discussion}
To summerize, the simple one-dimensional model, introduced here, exhibits a
first order phase transition. Our approach is a generalization of the mean field
approximation. One of the methods to perform the mean field approximation starts
by assuming a probability distribution of the product form, neglecting the
correlations between successive units of the polymer. The parameters of the
probability distribution are then varied to minimize the free energy of the
total chain. We improve on this scheme by starting with the probability
distribution of an exactly solvable model. This is an improvement because now
the correlations between successive units are not neglected. The probability
distribution depends on two model parameters $\epsilon$ and $\mu$. Next
$\epsilon$ and $\mu$ are varied to minimize the free energy of the total chain,
as calculated using the correlation functions of the exactly solvable model.

In this methodology we need two models. At one hand the exactly solvable model
whose correlation functions are used. At the other hand the more realistic
model, which is treated in the mean field approach. The exactly solvable model
is a random walk with a one step memory effect. This model is used in
\cite{ref11} to compare the outcome of two different single-molecule
experiments. In that paper the Hamiltonian is proportional with the number of
kinks. In the present paper, we increased the physical relevance by defining the
Hamiltonian by (\ref{energie}). In this way the first order phase transition,
which is seen in force-clamp experiments, is included in the theory. To increase
the physical relevance even further we should look for mean-field approximations
of realistic Hamiltonians. For example, in \cite{ref5}, numerical
simulations are performed with a Hamiltonian which contains a quartic potential
in combination with a Lennard-Jones potential. The analytic treatment
of such a realistic Hamiltonian poses technical problems
and it is not yet clear how to overcome these.

The model under study describes a one-dimensional finite system. Therefore,
no phase transition is expected. However, the present model does exhibit a
phase transition. The reason is the introduction of a mean-field-like interaction,
see the
definition (\ref{energie}) of the Hamiltonian. It is well known that such
interactions can cause phase transitions even in finite, one-dimensional
systems. For instance, in \cite{ref13,ref14} a one-dimensional, infinite
Ising model is considered and the Hamiltonian contains a long-range,
mean-field-like coupling. The canonical phase diagram is derived in
\cite{ref13} and exhibits a phase transition. In \cite{ref14} the model
is further analyzed within the microcanonical ensemble. The authors show that
the phase diagrams of the canonical and the microcanonical ensemble are not
equivalent for this model.

In force-clamp experiments, one is able to study the mechanical unfolding of
polymers in the fixed force ensemble. Both experiments \cite{ref1} and
numerical simulations \cite{ref2} show that a certain protein (called
ubiquitin) unfolds in a single step. This is in qualitative agreement with the
results of our one-dimensional model. In order to compare theory and experiment
quantitatively, our model has to be generalized to higher dimensions. This is a
technical complication. Our present model cannot be used to discuss the results
of experiments performed in the fixed stretch ensemble \cite{ref3,ref4}.
 The reason for this is that the important mean-field-like term of the present
model becomes zero in this ensemble. Our present model can also not produce an
unfolding process in multiple steps, which is observed in
\cite{ref2,ref3,ref4}. To describe these experiments, the excluded
volume effect has to be taken into account more accurately. One obtains then a
so called self-avoiding walk. In \cite{ref15} such self-avoiding walks are
studied numerically in the fixed force ensemble. The authors show that at low
temperatures the force-extension curve has a multi-step pattern due to finite
size effects.

\appendix
\section{}\label{app1}
In this appendix the results of two straightforward but tedious calculations are
given. The average of the squared end-to-end distance is obtained with the
generating function method. The result is
\begin{eqnarray}\label{gem_x_kwa}
\frac{\langle x^2\rangle}{n^2a^2}&=&\left(\frac{\langle
x\rangle}{na}\right)^2\cr
& &+\frac{4}{n^2}\frac{(1-\epsilon)(1-\mu)}{
(2-\epsilon-\mu)^4}\left[
n(\epsilon+\mu)(2-\epsilon-\mu)+2(\epsilon+\mu-1)\left((\epsilon+\mu-1)^n-1\right)\right].
\end{eqnarray}
The set of equations $\partial G/\partial\epsilon=0$, $\partial G/\partial\mu=0$
is solved for $\beta$ and $F$ for arbitrary $n$. This gives
\begin{eqnarray}
\label{bet_F_app}
\frac{a}{h}F&=&\frac{1}{2h\beta}\ln\frac{\epsilon}{\mu}-2\frac{\epsilon-\mu}{
(2-\epsilon-\mu)^2}\Bigg[2+\epsilon+\mu
+\frac{2}{n}\frac{1-2\epsilon-2\mu}{2-\epsilon-\mu}\cr
& &\qquad
+\frac{2}{n}\frac{(\epsilon+\mu-1)^n(2\epsilon+2\mu-1+n(2-\epsilon-\mu))}{2-\epsilon-\mu}\Bigg]
\cr
h\beta&=&\frac{n}{8}\frac{(2-\epsilon-\mu)^3}{n(2-\epsilon-\mu)\left[
1+(\epsilon+\mu-1)^n\right]-(\epsilon+\mu)\left[1-(\epsilon+\mu-1)^n\right]}
\ln\frac{1-\epsilon}{\epsilon}\frac{1-\mu}{\mu}.
\end{eqnarray}


\begin{thebibliography}{99}
\bibitem{ref1} M. Schlierf, H. Li, J. M. Fernandez, {\sl The unfolding
kinetics of ubiquitin captured with single-molecule force-clamp techniques,}
Proc. Natl. Acad. Sci. USA 101, 7299-7304 (2004)

\bibitem{ref2} P. Szymczak, M. Cieplak, {\sl Stretching of proteins in a
force-clamp,} J. Phys.:Condens. Matter 18, L21-L28 (2006)

\bibitem{ref3} C. G. Baumann, V. A. Bloomfield, S. B. Smith, C. Bustamante,
M. D. Wang, S. M. Block, {\sl Stretching of Single Collapsed DNA Molecules,}
Biophys. J. 78, 1965-1978 (2000)

\bibitem{ref4} Y. Murayama, Y. Sakamaki, M. Sano, {\sl Elastic Response of
Single DNA Molecules Exhibits a Reentrant Collapsing Transition,} Phys. Rev.
Lett. 90, 018102 (2003)

\bibitem {ref5} M. Cieplak, T. Xuan Hoang, M. O. Robbins,
{\sl Thermal Folding and Mechanical Unfolding Pathways of Protein Secondary
Structures,}
Proteins Struct. Funct. Genet. 49, 104-113 (2002).

\bibitem {ref6} P. G. De Gennes,
{\sl Coil-stretch transition of dilute flexible polymers under ultrahigh
velocity gradients,}
J. Chem. Phys. 60, 5030-5042 (1974).

\bibitem{ref7} J. Dubbeldam, F. Redig, {\sl Multilayer Markov Chains with
Applications to Polymers in Shear Flow,} J. Stat. Phys. 125, 225-243 (2006)

\bibitem{ref8} P. Szymczak, M. Cieplak, {\sl Stretching of Proteins in a
Uniform Flow,} J. Chem. Phys. 125, 164903 (2006)

\bibitem{ref9} A. Halperin, E. B. Zhulina, {\sl On the Deformation Behaviour
of Collapsed Polymers,} Europhys. Lett. 15, 417-421 (1991)

\bibitem  {ref10} E. Van der Straeten, J. Naudts,
{\sl A two-parameter random walk with approximate exponential probability
distribution,}
J. Phys. A: Math. Gen. 39, 7245-7256 (2006).

\bibitem {ref11} E. Van der Straeten, J. Naudts,
{\sl A one-dimensional model for theoretical analysis of single molecule
experiments,}
J. Phys. A: Math. Gen. 39, 5715-5726 (2006).

\bibitem{ref13} J. F. Nagle, {\sl Ising chain with competing interactions,}
Phys. Rev. A 2, 2124-2128 (1970)

\bibitem{ref14} D. Mukamel, S. Ruffo, N. Schreiber, {\sl Breaking of
ergodicity and long relaxation times in systems with ling-range interactions,}
Phys. Rev. Lett. 95, 260604 (2005)

\bibitem{ref15} D. Marenduzzo, A. Maritan, A. Rosa, F. Seno, {\sl Stepwise
unfolding of collapsed polymers,} Eur. Phys. J. E 15, 83-93 (2004)

\end{thebibliography}
\end{document}